\documentclass[11pt,prd,superscriptaddress,nofootinbib]{revtex4}
\usepackage[usenames,dvipsnames]{xcolor}
\def\be{\begin{equation}}
\def\ee{\end{equation}}
\def\ba{\begin{array}}
\def\ea{\end{array}}
\def\bea{\begin{eqnarray}}
\def\eea{\end{eqnarray}}

\def\pa{\partial}

\begin{document}

\begin{center}
		{\Large \bf $\pi$ meson-octet baryon coupling constant obtained by hard-wall AdS/QCD model }
		
		\vskip 1. cm
		{Shahin Mamedov $^{a,c,e,f}$\footnote{corresponding author: sh.mamedov62@gmail.com }} and
		{Shahnaz Taghiyeva $^{b,d,e}$\footnote{shahnazilgarzade@gmail.com}},
		
		\vskip 0.5cm
		
		{\it $^a\,$ Institute for Physical Problems, Baku State University,
			
			Z. Khalilov street 23, Baku, AZ-1148, Azerbaijan,}
		\\ \it \indent $^b$Theoretical Physics Department, Physics Faculty, Baku State University,
		
		Z.Khalilov Street 23, Baku, AZ-1148, Azerbaijan,
		\\ \it \indent $^c$Institute of Physics, Ministry of Science and Education,
				
		H. Javid avenue 33, Baku, AZ-1143,
		Azerbaijan,\\ \it \indent $^d$Shamakhy Astrophysical Observatory, Ministry of Science and Education,
		
		H.Javid 115, Baku, AZ 1000, Azerbaijan,		
		\\ \it \indent $^e$ Center for Theoretical Physics, Khazar University,
		
		41 Mehseti Street, Baku, AZ-1096, Azerbaijan,
  \\ \it \indent $^f$ Lab. for Theor. Cosmology, International Centre of Gravity and Cosmos, Tomsk State University of Control Systems and Radio Electronics (TUSUR),
  
  634050 Tomsk, Russia \\

	\end{center}

\centerline{\bf Abstract} \vskip 4mm
We investigate a strong coupling constant between  $\pi$ meson and octet baryon using the hard-wall AdS/QCD model, where $SU(3)_{f}$ symmetry-breaking effects are considered. Also, we have mentioned briefly the determining of the profile functions for pseudoscalar meson and baryon fields. The values of couplings obtained here agree with the ones obtained within other theoretical models.
\vspace{1cm}

\section{Introduction}

The anti-de-Sitter/Conformal Field Theory (AdS/CFT) correspondence between two theories is strong-weak, meaning that it relates the strong coupling sector of one theory to the weak coupling sector of the other and vice versa. The AdS/QCD (Anti de-Sitter/Quantum ChromoDynamics) is a simple phenomenological model in five-dimensional curved spacetime, which describes meson and baryon sectors of QCD effectively. In Ref. \cite{zf} authors calculated the mass spectra of the baryon octet and their low-lying excited states. They briefly review the holographic model for spin $1/2$ baryons. A detailed description of meson-nucleon coupling within the hard-wall AdS/QCD model was given in Ref. \cite{mn}. In Ref.  \cite{hi} a hard-wall model for low-lying spin $1/2$ baryons that includes the effect of chiral symmetry breaking was proposed to explain the parity-doublet pattern of excited baryons. 
 The baryon-meson coupling constants were calculated for the spin $1/2$  baryon octet and spin $3/2$ decuplet in a unified approach relying on symmetry arguments such as the fact that the Yukawa couplings in Ref.  \cite{lk}. 

In Ref. \cite{ag}  the effects of flavor symmetry breaking on three-meson couplings and form factor were investigated and $g_{\rho^{n}\pi \pi}$, $g_{\rho^{n}\rho \rho}$, $g_{\rho^{n} KK}$, $g_{\rho^{n} K^{*}K^{*}}$, $g_{\rho^{n} DD}$ and $g_{\rho^{n} D^{*}D^{*}}$  strong couplings were predicted using hard-wall holographic model.  The two-flavor hard-wall holographic model was extended to four flavors in Ref. \cite{sm} and the strong couplings of $( D^{(*)}, D, A)$, $( D^{(*)}, D^{(*)}, V ) $, $ (D_{1}, D_{1}, P )$, $(\Psi, D^{(*)}, D, P)$, and $(\Psi, D^{(*)}, D,A)$  were calculated after analyzing the wave functions.

We obtain the coupling constants between light pseudoscalar meson $\pi$, and octet baryon using the hard-wall AdS/QCD model. The hard-wall model used a sharp cut-off in the extra dimension to realize the confinement and chiral symmetry breaking. In the hard wall model of AdS/QCD  the spacetime is given by a five-dimensional anti-de Sitter  metric with the additional dimensional labelled as $z$:
\begin{equation}
ds^2=g_{MN}dx^{M}dx^{N}=\frac{1}{z^2}(\eta_{\mu\nu}dx^{\mu}dx^{\nu}-dz^{2}), 
\label{1}
\end{equation}
where $\eta_{\mu\nu}=diag(1,-1,-1,-1)$, $\mu,\nu=0,1,2...d-1$;  $g=|det(g_{MN})|=1/z^{10}$ the magnitude of the determinant of $g_{MN}$. Thus, the model is defined in the range: $\epsilon\leq z \leq z_{m}$,   $\epsilon$ and $z_{m}$ are the inverses of the ultraviolet (UV) and infrared (IR) cutoff scales, respectively, in the hard-wall model.

The paper is structured as follows: In the next section, we briefly review our approach-profile functions in the hard-wall model for pseudoscalar and baryon fields. In Sec. III, we define the between $\pi$ meson- baryon coupling constant in the hard-wall model. Finally, in Sec. IV, we discuss our results and compare our final results to existing theoretical results and experimental data.

\section{Profile functions in the hard wall}

\subsection {The wave function of the pseudoscalar meson}
Bulk action for the meson sector, which includes the pseudoscalar $X$ field and $A_{L,R}$ gauge fields, is written as below \cite{ek}:
\begin{equation}
S=\int d^{5}x \sqrt{g} Tr\left [ \left | DX \right |^{2}+3\left | X \right |^{2}-\frac{1}{4g_{5}^{2}}\left (F_{L}^{2}+F_{R}^{2}  \right ) \right ], \label{14}
\end{equation}
where $DX$  is the covariant derivative of the bulk scalar field: $D^{M}X=\partial^{M}X-iA_{L}^{M}X+iXA_{R}^{M}$ and $F_{L,R}^{MN}$ is the  chiral gauge field strength: $F_{L,R}^{MN}=\partial^{M}A_{L,R}^{N}-\partial^{N}A_{L,R}^{M}-i\left [ A_{L,R}^{M}, A_{L,R}^{N} \right ]$. The equation of motion for the pseudoscalar meson is derived using from the least action principle and $A_{5}=0$ gauge condition in this gauge \cite{zf}:

\begin{equation}
\partial_{z}\left ( \frac{\partial_{z}P_{\phi}}{z} \right )+\frac{4g_{5}^{2}v_{u}^{2}}{z^{3}}\left ( f_{\pi}-P_{\phi} \right )=0,  
\label{15}
\end{equation}

\begin{equation}
 -q^{2}\pa_{z}P_{\phi} +\frac{4g_{5}^{2}v_{u}^{2}}{z^{2}}\pa_{z}f_{\pi}=0, 
 \label{16}
\end{equation}
where $P_{\phi}(z)$ comes from that of the radial component of the axial-vector field, $f_{\pi}$ comes from the KK decomposition of the bulk pion field. As in the calculation of baryon spectrum, we only consider the $m_{u,d} = 0$ case in which the pion as Nambu-Goldstone boson has zero mass $ (q^{2} = 0)$, so we have $f^{'}_{\pi} (z) = 0$, which also indicates with the normalization condition $f_{\pi}(z\rightarrow 0)=0$ that 
$ f_{\pi}(z) = 0$. Then the equation of motion (\ref{16}) can be simplified as following:

\begin{equation}
z\partial_{z}\left ( \frac{\partial_{z} P_{\phi}}{z}\right )-\frac{g_{5}^{2}v_{u}^{2}}{z^{2}}P_{\phi}=0.
\label{17}
\end{equation}
The solution of the  equation (\ref{17}) is found  as below \cite{tensor}:
\begin{equation}
P_{\phi}(z)=N \xi z^{3} \left [I_{2/3} \left ( \frac{\sqrt{2}}{3} \xi z^{3} \right )-\frac{I_{2/3}\left ( \frac{\sqrt{2}}{3}\xi z^{3}_{m} \right )}{K_{2/3} \left ( \frac{\sqrt{2}}{3} \xi z^{3}_{m}\right) K_{2/3}\left (\frac{\sqrt{2}}{3} \xi z^{3} \right)} \right].
\label{18}
\end{equation}
Here $\xi=3.77 z_{m}^{-3}$ , $I_{2/3}$, $K_{2/3}$ are Bessel functions. The normalization constant  $N$ is determined from the normalization condition \cite{FizRev, tensor2023}, and equals:
\begin{equation}
N^{-1/2}=\frac{I_{2/3}\left ( \frac{\sqrt{2}}{3}\xi z^{3}_{m}\right)}{2K_{2/3}}\left ( \frac{\sqrt{2}}{3}\xi z^{3}_{m}\right)^{2} \left [ \sqrt{3} \pi I_{2/3}\left ( \frac{\sqrt{2}}{3}\xi z^{3}_{m}\right) +3K_{2/3} \left ( \frac{\sqrt{2}}{3}\xi z^{3}_{m}\right) \right ].  \label{19}
\end{equation}

\subsection{Profile function for baryon field}

Let us briefly present obtaining the profile function for the baryon field within the hard-wall model. The bulk Lagrangian for the fermion field, which describes the baryons on the boundary theory, is written as below:
\begin{eqnarray}
L_{Baryon}=\sqrt{-g}\left [\frac{i}{2}\overline{N}_{1}e^{M}_{A}\Gamma^{A}\nabla_{M}N_{1}-\frac{i}{2}(\nabla_{M}^{\dagger}\overline{N}_{1})e^{M}_{A}\Gamma^{A}N_{1}-m_{5}\overline{N}_{1}N_{1}\right], 
 \label{21}
\end{eqnarray}
where $e^{M}_{A}=\frac{1}{z}\eta_{M}^{A}$ is the vielbein satisfying $g_{MN}=e^{M}_{A}e^{N}_{B}\eta_{AB}$, $\Gamma^{A}=(\gamma^{\mu},-i\gamma^{5})$ are the Dirac matrices and satisfy: $[\Gamma^{A},\Gamma^{B}]=2\eta^{AB}$ and $\nabla_{M}$ is the gauge and Lorentz covariant derivative:
\begin{equation}
\nabla_{M}=\partial_{M}+\frac{i}{4}\omega_{M}^{AB}\Gamma_{AB}-i(A_{L}^{a})_{M}t^{a}.
 \label{22}
\end{equation}
Here $\Gamma^{AB}=\frac{1}{2i}[\Gamma^{A},\Gamma^{B}]$,  $\omega_{M}^{AB}$ is spin connection with nonzero components:  $\omega_{M}^{5A}=-\omega_{M}^{A5}=\frac{1}{z}\delta_{M}^{A}$.
The Yukawa interaction in the bulk is introduced to include the chiral symmetry breaking:
\begin{equation}
L_{Yukawa}=-g_{Y}\overline{N}_{1}XN_{1}-g_{Y}\overline{N}_{2}X^{\dagger}N_{2},
 \label{23}
\end{equation}
where $g_{Y}$ is the  Yukawa constant. The bulk baryon field can be written the chiral components:
\begin{equation}
N_{1}=N_{1L}+N_{1R} , 
\label{24}
\end{equation}
where $i\Gamma^{5}N_{1L}=N_{1L}$ and $i\Gamma^{5}N_{1R}=-N_{1R}$.
Then a Kaluza-Klein decomposition for $N_{1L,R}$ is performed to yield the following form:
\begin{equation}
N_{1L,R}(x,z)=\sum_{n}\int \frac{d^{4}p}{(2\pi)^{4}} f_{1L,R}(z)\psi_{L,R}^{(n)}(p)e^{-ipx},
\label{25}
\end{equation}
\begin{equation}
\not{\!}{p}\Psi_{(1L,R)}(p)=|p|\Psi_{(1R,L)}(p).
\label{26}
\end{equation}
We get an equation for $f_{L,R}(z)$:

\begin{eqnarray}
\ 
\left(\begin{array}{cc}
        \pa_{z}-\frac{\Delta}{z}&-\frac{g_{Y}v(z)}{z}\\
        \frac{g_{Y}v(z)}{z}^{\dagger}&\partial_{z}-\frac{4-\Delta}{z}\\
      \end{array}\right)
      \ 
\,
\ 
      \left  
(\begin{array}{c}
              f_{1L}\\
              f_{2L}\\
            \end{array}\right)
            \ 
            & = &
m\,
\
            \left(\begin{array}{c}
                    f_{1R}\\
                    f_{2R}\\
                  \end{array}\right),
\ 
\end{eqnarray} \label{27}
where $v(z)$ has the following form for octet baryon:
\begin{equation}
v_{p,n}(z)=\left (3c_{2}-c_{1}\right)v_{u}(z),
\label{28}
\end{equation}
\begin{equation}
v_{\Sigma s}(z)=2c_{2}v_{u}(z)+\left (c_{2}-c_{1}\right)v_{s}(z),
\label{30}
\end{equation}
\begin{equation}
v_{\Xi s}(z)=\left (c_{2}-c_{1}\right)v_{u}(z)+2c_{2}v_{s}(z).
\label{31}
\end{equation}
In the chiral limit, the $f_{1}$ and $f_{2}$ sectors are completely decoupled from each other and the equation of motion for $f_{1}$ sector is found as follows \cite{mn}:
\begin{equation}
\left(\partial_{z}^{2}-\frac{4}{z}+\frac{6+m_{5}-m_{5}^{2}}{z^{2}}\right)f_{1L}^{n}(z)=-m_{n}^{2}f_{1L}^{n}(z),
 \label{32}
\end{equation}
\begin{equation}
\left(\partial_{z}^{2}-\frac{4}{z}-\frac{6+m_{5}-m_{5}^{2}}{z^{2}}\right)f_{1R}^{n}(z)=-m_{n}^{2}f_{1R}^{n}(z),
 \label{33}
\end{equation}
where $m_{n}$ represents the KK masses for the baryons. The solutions of the  (\ref{32}),  (\ref{33})  equations can be obtained as:
\begin{equation}
f_{1L,R}^{n}(z)=z^{\frac{5}{2}}\left[c_{1}J_{|m_{5}\mp\frac{1}{2}|}(m_{n}z)+c_{2}Y_{|m_{5}\mp\frac{1}{2}|}(m_{n}z)\right].
\label{34}
\end{equation}
The scaling dimension of the baryon operator is $\Delta=\frac{9}{2}$, according to $m_{5}^{2}=(\Delta-2)^{2}$ the of $m_{5}$ value will be $m_{5}=\pm\frac{5}{2}$. Therefore, we can write  $f_{L,R}$ as follows:
\begin{equation}
f_{1L}^{n}=z^{\frac{5}{2}}\left [c_{1L}J_{2}(m_{n}z)+c_{2L}Y_{2}(m_{n}z) \right], 
\label{35}
\end{equation}
\begin{equation}
f_{1R}^{n}=z^{\frac{5}{2}}\left [c_{1R}J_{3}(m_{n}z)+c_{2R}Y_{3}(m_{n}z) \right].
\label{36}
\end{equation}
$c_{2}$ constant is found from  the UV  boundary condition $f_{1L}^{n}(0)=0$ we get:
\begin{equation}
c_{2L}^{n}=-\frac{J_{2}(m_{n}\varepsilon)}{Y_{2}(m_{n}\varepsilon)}c_{1L}^{n}=\frac{\pi}{2}\left(\frac{m_{n}\varepsilon}{2}\right)^{4}c_{1L}^{n}\rightarrow0. 
\label{37}
\end{equation}
Finally, $f_{1L}$ accepts simple form:
\begin{equation}
f_{1L}^{n}=c_{1L}z^{\frac{5}{2}}J_{2}(m_{n}z).
\label{38}
\end{equation}
 $f_{1R}$ is found by taking into account   $f_{1L}$ in (12):
\begin{equation}
f_{1R}^{n}=c_{1R}z^{\frac{5}{2}}J_{3}(m_{n}z),
\label{39}
\end{equation}
 $f_{2R,L}$  is found applying $f_{1R}^{n}(0)=0$ boundary condition and taking  $m_{5}=-\frac{5}{2}$ :
\begin{equation}
f_{2R}^{n}=c_{2R}z^{\frac{5}{2}}J_{2}(m_{n}z),
\label{40}
\end{equation}
\begin{equation}
f_{2L}^{n}=-c_{2L}z^{\frac{5}{2}}J_{3}(m_{n}z).
\label{41}
\end{equation}
$c_{1,2}$ constants are found from the normalization condition \cite{mn}:
\begin{equation}
|c_{1,2}|=\frac{\sqrt{2}}{z_{m}J_{2}(m_{n}z_{m})}.
\label{42}
\end{equation}

\section{$\pi$ meson-octet baryon coupling constant}

The action for the interaction between bulk fields in the AdS space  can be written as the $5D$  integral:
\begin{equation}
S=\int_{0}^{z_{m}}d^{4}xdz\sqrt{g}L_{int}(x,z).
\label{43}
\end{equation}
Here $L_{int}$ is the interaction Lagrangian. The interaction Lagrangian for the $\pi$ meson and octet baryon is given as:
\begin{equation}
L_{\pi bb}=-g_{Y}\left[\overline{B}_{1}XB_{2}+\overline{B}_{2}X^{+}B_{1}\right].
\label{47}
\end{equation}
Here $X$ is a scalar field and is defined as $ X=v(z)e^{iPx}$.   $B_{1,2}$ baryon fields can be written as follows by Fourier transformation:
\begin{equation}
B_{1,2L,R}=\frac{1}{(2\pi)^{4}}\int d^{4}p f_{1,2L,R} \Psi(p)e^{-ipx}.
\label{48}
\end{equation}
The map between bulk and boundary quantities is called GKPW (Gubser-Klebanov-Polyakov-Witten) formula which states that:
\begin{equation}
Z_{sugra}|J|\equiv \int_{\Phi \sim J}D\Phi e^{-S[\Phi]}=\langle e^{-\int_{\partial AAdS}JO}\rangle\equiv Z_{QFT},
\label{49}
\end{equation}
where the left-hand side is the supergravity partition function in anti-de Sitter space, and the right-hand side is the partition function of the Quantum Field Theory (QFT) boundary of the asymptotically this spacetime. As seen from (\ref{49}) relation, $\Phi \sim J$ means that the field theory side $J$  is a source coupled to an operator $O$ which is dual to $\Phi$. In the limit of small string coupling and large string tension, the equivalence between the partition function of the QFT and the partition function of supergravity will be as follows:
\begin{equation}
e^{-S_{on-shell}[J]}=\langle e^{-\int_{\partial AAdS}JO}\rangle_{N\rightarrow \infty, \lambda\rightarrow \infty}.
\label{50}
\end{equation}
Here, the left-hand side is on-shell supergravity action, and the right-hand side is the QFT partition function in the large $ N$  and strong coupling limits. We can define expectation values of CFT observables by taking functional derivatives of this relation
\begin{equation}
\langle O(x)\rangle \sim \frac{\delta S_{on-shell}}{\delta \Phi(x)}|_{\Phi(x)=0} .
\label{51}
\end{equation}
According to AdS/CFT correspondence the relation  (\ref{51}) can be written  for the vacuum expectation value of the baryon current $ J_{\mu}$:
\begin{equation}
\langle J_{\mu}\rangle=-i\frac{\delta Z_{AdS}}{\delta \tilde{V}_{\mu}^{0}}|_{\tilde{V}_{\mu}^{0}}=0,
\label{52}
\end{equation}
where $\tilde{V}_{\mu}^{0}=\tilde{V}_{\mu}(q,z=0)=V(q)$ is the UV boundary value of the vector field $(V(z=0)=1)$. On the other side, a current  in the QCD theory  is defined as:
\begin{equation}
J_{5}(p^{'},p)=g_{\pi bb}\overline{u}(p^{'})\gamma_{5} u(p). 
\label{53}
\end{equation}
The $\pi$ meson - octet baryon coupling constant $g_{\pi bb}$ defines from a comparison of (\ref{52}) and (\ref{53}) equations:
\begin{equation}
g_{\pi bb}=\int^{z_{m}}_{0}\frac{dz}{z^{5}}v(z)g_{Y}P(z)\left ( f_{1R}(z)f_{2L}(z)-f_{1L}(z)f_{2R}(z) \right ).
\label{57}
\end{equation}

\section{Numerical Results}
Having fixed the parameters $\sigma$ and $m_q$ the constant $g_{\pi bb}$ (\ref{57}) can be calculated numerically. We apply the $\sigma = (0.213)^{3}$ $GeV^{3}$ and $m_{q} = 0.0083$ $GeV$ values for these parameters, which were fixed in Ref. \cite{Cherman}. For the constant $g_{Y}$  we take the value $g_{Y} =9.182$ which was established in Ref. \cite{Hong}. In Table 1 we compare our results to the ones obtained within the chiral soliton model and light cone QCD sum rule. 

As seen from the Table 1, $g_{\pi \Sigma^{0} \Sigma^{0}}$ is close to the chiral soliton model result. The numerical value for the between $ \pi$ meson and $\Xi^{0} $ baryon coupling constant corresponds to the result obtained within the light cone QCD sum rules framework. For the $g_{\pi \Lambda \Lambda}$ coupling constant hard-wall result is very small $ (g_{\pi \Lambda \Lambda}=4,59652 \times 10^{-6} ) $. The result here is reasonable from a physical
interpretation point of view. Since the current of the $\Lambda$ baryon is antisymmetric concerning the $u$, $d$ quarks, and we use a chiral limit state, this coupling constant is automatically equal to zero.

Table 1. $\pi$ meson-octet baryon coupling constant

\begin{center}

\begin{tabular}{|c|c|c|c|c|c|c|c|}
  \hline
  Model & $ g_{\pi NN}$ & $g_{\pi \Sigma^{0} \Sigma^{0} }$ & $ g_{ \pi \Sigma^{-} \Sigma^{-}}$ & $g_{\pi \Sigma^{+} \Sigma^{+} } $ & $ g_{\pi \Xi^{0} \Xi^{0}}$ & $g_{\pi \Xi^{-} \Xi^{-}}$  \\
  [0.5ex]
  \hline\hline
  Hard-wall model & 2.941 &3.516 & 3.554 & 3.478 & 2.4799 & 2.516\\
  \hline
  Chiral soliton model \cite{yang} & $3.524 \pm 0.012$ & $3.356 \pm 0.014$ & -&-& $-1.24 \pm 0.009$ & - \\
  \hline
  Light cone QCD sum rule \cite{Aliev} & - &  $2.8 \pm 0.3$ & - & - & $2.4 \pm 0.2$ & - \\
  \hline
\end {tabular}

\end{center}

\bibliographystyle{IEEEtran}
\bibliography{sources.bib}

\begin{thebibliography}{10}
\providecommand{\url}[1]{#1}
\csname url@samestyle\endcsname
\providecommand{\newblock}{\relax}
\providecommand{\bibinfo}[2]{#2}
\providecommand{\BIBentrySTDinterwordspacing}{\spaceskip=0pt\relax}
\providecommand{\BIBentryALTinterwordstretchfactor}{4}
\providecommand{\BIBentryALTinterwordspacing}{\spaceskip=\fontdimen2\font plus
\BIBentryALTinterwordstretchfactor\fontdimen3\font minus \fontdimen4\font\relax}
\providecommand{\BIBforeignlanguage}[2]{{%
\expandafter\ifx\csname l@#1\endcsname\relax
\typeout{** WARNING: IEEEtran.bst: No hyphenation pattern has been}%
\typeout{** loaded for the language `#1'. Using the pattern for}%
\typeout{** the default language instead.}%
\else
\language=\csname l@#1\endcsname
\fi
#2}}
\providecommand{\BIBdecl}{\relax}
\BIBdecl

\bibitem{zf}
Z.~Fang, ``Holographic model for the baryon octet,'' \emph{Physical Review D}, vol.~94, no.~7, p. 074017, 2016.

\bibitem{mn}
N.~Maru and M.~Tachibana, ``Meson--nucleon coupling from ads/qcd,'' \emph{The European Physical Journal C}, vol.~63, pp. 123--132, 2009.

\bibitem{hi}
D.~K. Hong, T.~Inami, and H.-U. Yee, ``Baryons in ads/qcd,'' \emph{Physics Letters B}, vol. 646, no.~4, pp. 165--171, 2007.

\bibitem{lk}
L.~L. Lopes, K.~D. Marquez, and D.~P. Menezes, ``Baryon coupling scheme in a unified su (3) and su (6) symmetry formalism,'' \emph{Physical Review D}, vol. 107, no.~3, p. 036011, 2023.

\bibitem{ag}
A.~Ballon-Bayona, G.~Krein, and C.~Miller, ``Strong couplings and form factors of charmed mesons in holographic qcd,'' \emph{Physical Review D}, vol.~96, no.~1, p. 014017, 2017.

\bibitem{sm}
S.~Momeni and M.~Saghebfar, ``Erratum to: Charm meson couplings in hard-wall holographic qcd,'' \emph{The European Physical Journal C}, vol.~82, no.~8, p. 709, 2022.

\bibitem{ek}
J.~Erlich, E.~Katz, D.~T. Son, and M.~A. Stephanov, ``Qcd and a holographic model of hadrons,'' \emph{Physical Review Letters}, vol.~95, no.~26, p. 261602, 2005.

\bibitem{tensor}
E.~Katz, A.~Lewandowski, and M.~D. Schwartz, ``Tensor mesons in ads/qcd,'' \emph{Physical Review D}, vol.~74, no.~8, p. 086004, 2006.

\bibitem{FizRev}
------, ``Tensor mesons in ads/qcd,'' \emph{Physical Review D}, vol.~74, no.~8, p. 086004, 2006.

\bibitem{tensor2023}
S.~Mamedov, Z.~Hashimli, and S.~Jafarzade, ``Tensor meson couplings in ads/qcd,'' \emph{Physical Review D}, vol. 108, no.~11, p. 114032, 2023.

\bibitem{Cherman}
\BIBentryALTinterwordspacing
A.~Cherman, T.~D. Cohen, and E.~S. Werbos, ``Chiral condensate in holographic models of qcd,'' \emph{Physical Review C}, vol.~79, no.~4, Apr. 2009. [Online]. Available: \url{http://dx.doi.org/10.1103/PhysRevC.79.045203}
\BIBentrySTDinterwordspacing

\bibitem{Hong}
\BIBentryALTinterwordspacing
D.~K. Hong, H.-C. Kim, S.~Siwach, and H.-U. Yee, ``The electric dipole moment of the nucleons in holographic qcd,'' \emph{Journal of High Energy Physics}, vol. 2007, no.~11, p. 036–036, Nov. 2007. [Online]. Available: \url{http://dx.doi.org/10.1088/1126-6708/2007/11/036}
\BIBentrySTDinterwordspacing

\bibitem{yang}
G.-S. Yang and H.-C. Kim, ``Meson--baryon coupling constants of the su (3) baryons with flavor su (3) symmetry breaking,'' \emph{Physics Letters B}, vol. 785, pp. 434--440, 2018.

\bibitem{Aliev}
T.~M. Aliev, K.~Azizi, and M.~Savc{\i}, ``Strong transitions of decuplet to octet baryons and pseudoscalar mesons,'' \emph{Nuclear Physics A}, vol. 847, no. 1-2, pp. 101--117, 2010.

\end{thebibliography}

\end{document}